\documentclass[aps,superscriptaddress,twocolumn,twoside,floatfix,prl,a4paper]{revtex4-2}
\usepackage{times}
\usepackage{epsfig}
\usepackage{amsfonts}
\usepackage{amsmath}
\usepackage{amssymb,amsthm}
\usepackage{color}
\usepackage{multirow}
\usepackage{braket}
\usepackage{bbm}
\usepackage{latexsym}
\usepackage{amsfonts}
\usepackage{mathrsfs}
\usepackage{natbib}
\usepackage{verbatim}
\usepackage{gensymb}
\usepackage{graphicx}
\allowdisplaybreaks

\usepackage[colorlinks=true,linkcolor=blue,citecolor=magenta,urlcolor=blue]{hyperref}
\allowdisplaybreaks

\begin{document}

\title{Effective Repulsive Action of Gravitational Quantum Superpositions Under Postselection}

\author{Sougato Bose}
\affiliation{Department of Physics and Astronomy, University College London, Gower Street, London WC1E 6BT, England, United Kingdom}
\author{Lev Vaidman}\affiliation{School of Physics and Astronomy, Tel-Aviv University, Tel-Aviv 69978, Israel}


\begin{abstract}
A classic feature of gravity is that it is an attractive force. If a source mass is prepared in a localized (classical-like) state, it will cause another probe mass to move towards it. Here we consider the situation in which a source mass is prepared in a quantum superposition of distinct spatial states while a probe mass interacts with it. Conditional on the detection of the source mass in a specific state, the probe mass will be found to move away from the source mass (repulsion). This signifies the quantum superposition of gravitational forces acting on the probe mass and thereby the fact that spacetime can exist in quantum superpositions. The technique used is the repulsive effect arising from an anomalous negative weak value. A potential experimental implementation with spin bearing nanocrystals is outlined.
\end{abstract}
\maketitle

{\em Introduction:} 
Quantum mechanics is extremely successful. Classical General Relativity is successful too, but currently there
is no accepted theory of quantum gravity. Indeed there are widely studied theories in which gravity is treated quantum mechanically \cite{zwiebach2004first,rovelli2015covariant,oriti2009approaches}, growing out of historical efforts \cite{Bronstein2012republication,gupta1952quantization,feynman1963quantum,dewitt1967quantum,weinberg1965photons} which have a low energy limit \cite{donoghue1995introduction} consistent with our current observations. However, to date, there are no uncontroversial {\em empirical} proofs of the quantum nature of gravity or equivalently, the quantum nature of spacetime, as gravity is nothing but the curvature of spacetime. ``Whether" gravity is quantum is currently an open question \cite{carlip2008quantum} including models which combine quantum matter and forces with  classical spacetime, introducing stochasticity for consistency \cite{oppenheim2022constraints}. One must clarify here what an unambiguous proof of gravity being qualitatively quantum would be: it should be an effect that arises from spacetime obeying the fundamental principles of quantum mechanics such as the quantum superposition principle. In this context, thus, the historically remarkable neutron interferometry experiments \cite{neutrongravity} and recent atom interferometer experiments \cite{mcguirk2002sensitive} can be explained by quantum systems evolving in a classical gravitational field, and thus {\em do not} tell us whether gravity is quantum. For much of the community, there is no serious doubt that such superpositions exist: gedanken experiments have been proposed in the past, e.g. superposition of evolutions with different gravitational potentials \cite{Timemachine}. Recently, the mediation of entanglement via gravity has been suggested as a feasible empirical test of its quantum nature \cite{bose2016matter,bose2017spin,marletto2017gravitationally,biswas2022gravitational,marshman2020locality,christodoulou2019possibility,bose2022mechanism,christodoulou2022locally,carney2022newton,carney2021using,galley2022no,qvarfort2020mesoscopic,krisnanda2020observable,chevalier2020witnessing,feng2022amplification,schut2023micron,yi2021massive,kaku2023enhancement,braccini2026mass,bose2025massive,marletto2025quantum,bose2025spin}. Some other approaches have also been suggested \cite{hanif2024testing,haine2021searching,howl2021non,lami2024testing,tobar2024detecting}.

\begin{figure}
    \centering
    \includegraphics[scale=0.47]{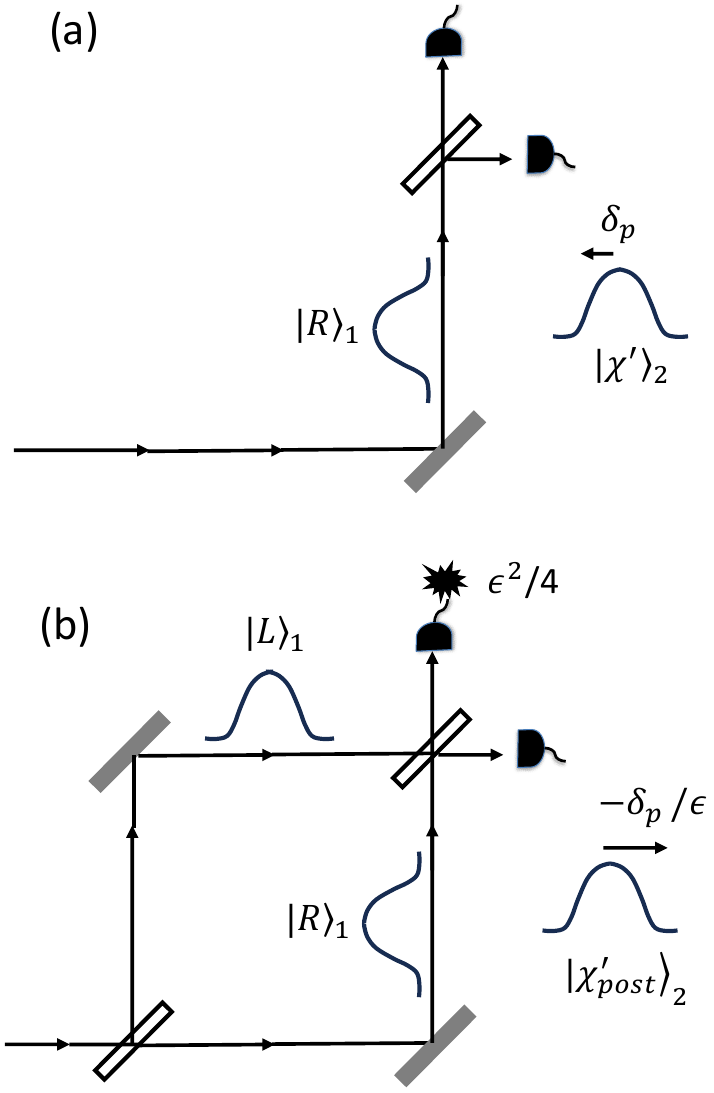}
    \caption{Schematic idea of the repulsive effect of a quantum superposition of gravitational fields under postselection. (a) the usual gravitational attractive effect of mass-1 in a localized state $|R\rangle_1$ on mass-2 in an initial gaussian (localized) state. This attraction leads to mass-2 gaining a momentum $\delta_p$ towards mass-1. (b)  mass-1 in a superposition of two localized states $|L\rangle_1$ and $|R\rangle_1$ (given by Eq.(\ref{initial})) and further subjected to postselection (given by Eq.(\ref{final})). Under postselection, mass-2 appears to acquire a momentum $\delta_p/\epsilon$ away from mass-1.}
    \label{repulsivefig}
\end{figure}

However, we do not yet have decisive experimental results on the quantumness of gravity: this motivates us to continue to look for new, {\em qualitatively different}, tests. One of the successful quantum methods for measuring tiny effects is anomalous weak value amplification scheme \cite{AAV}. It allowed measuring the tiny Hall effect for light for the first time \cite{HK}. However, most importantly, weak value amplification is {\em itself} a quantum effect resulting from pre and post selection of quantum superposition states. What we propose here is to measure effect of gravitational coupling between two masses: one of which is in a superposition of two arms of an interferometer pre and postselected in such a way that we obtain anomalous weak value of the projection on the arm near the second mass. The observation of anomalous behavior will be a clear signature of quantumness of gravitational interaction. The classical mixture of interactions cannot be larger than any one of them. Here we will show that superposition of attraction and no interaction
will lead to repulsion which is even stronger than the attraction of the two masses.  

{\em Outline Scheme, Fig.1:} We prepare one mass (mass-1, which acts as the source mass) in a quantum superposition of two localized states, $|L\rangle_1$ and   $|R\rangle_1$ with the other mass (mass-2, which acts as the probe mass) in a localized state $|\chi\rangle_2 $. Temporarily (just to explain), we will {\em assume} a Mach-Zhender interferometer for mass-1, and later discuss the implementation. Mass-1 passes through an interferometer such that when it is detected in the low intensity (dark) port (as shown in Fig.\ref{repulsivefig}(b)), it gets an anomalous presence in the right arm which is near the second mass. The weak value of projection on this arm will have a large and negative value, so we expect amplification of the (weak) interaction and changing it from attraction to repulsion \cite{cenni2019effective,reznik2025interactions}. To this end, we arrange  the  state of the mass-1 preselected in state
\begin{equation}
    |\psi\rangle_1= \frac{1}{\sqrt{2}}\left(\sqrt{1+\epsilon}|L\rangle_1+\sqrt{1-\epsilon}|R\rangle_1\right),
\label{initial}
\end{equation}
where $\epsilon <<1$, and postselection of  state:
\begin{equation}
  |\phi\rangle_1=   \frac{1}{\sqrt{2}}\left(|L\rangle_1 -|R\rangle_1\right).
\label{final}
\end{equation}
The weak value of projection on the right arm (neglecting the interactions between pre and postselction) is \cite{AAV}
\begin{equation}
({\bf P}_R)_w=     \frac{\langle \phi| {\bf P}_R |\psi\rangle}{\langle \phi   |\psi\rangle}\approx - \frac{1}{\epsilon},
\end{equation}
The anomalously large and negative weak value suggests that our experiment will provide a desired result, but let us show this using standard quantum calculations.

{\em Explicit Calculations:} The superposition of different gravitational forces from mass-1  will act on mass-2.
The impulse of force on mass-2 due to the interaction is equal in size to the impulse obtained by mass-2, but the interferometer of mass-1 is arranged with non-Gravitational (electromagnetic)  strong interaction such that the gravitational effect on mass-1 can be neglected or properly taken into account.
Mass-2 is assumed to be freely moving under gravitational interaction.

 Let us begin in a situation where mass-2 is initially trapped in a harmonic well and has been cooled to its ground state centered at coordinate $x=0$ and momentum  $p=0$ and given by
\begin{equation}
|\chi\rangle_2 =\mathcal{N}\int_{-\infty}^{+\infty} e^{-\frac{x^2}{2\sigma^2}}|x\rangle dx =\mathcal{N}\int_{-\infty}^{+\infty} e^{-\frac{p^2}{2\Delta_p^2}}|p\rangle dp.
\end{equation}
where $\Delta_p=\frac{\hbar}{\sigma}$ is uncertainty in momentum, and $\mathcal{N}$ stands (here and in the rest of the paper) as the general symbol for normalizer relevant to different states.
Due to gravitational {\em attraction} from the $|R\rangle_1$ component, it gets a momentum $\delta_p$  towards mass-1 so that its state becomes
\begin{equation}
|\chi^{'}\rangle_2 = \mathcal{N}\int_{-\infty}^{+\infty} e^{-\frac{(p+\delta_p)^2}{2\Delta_p^2}}|p\rangle dp.
\end{equation}
Gravitational interaction is very small. The shift in the momentum $\delta_p$ is much smaller than momentum uncertainty $\Delta_p$, so we can decompose $|\chi'\rangle_2$ as a superposition of  $|\chi\rangle_2$ 
 and $|\chi^{\perp}\rangle_2$ such that  $_2\langle \chi^{\perp}|\chi\rangle_2=0$.
\begin{equation}
|\chi'\rangle_2 \approx ( |\chi\rangle_2- \frac{\delta_p}{\sqrt 2 \Delta_p}|\chi^{\perp}\rangle_2),
\end{equation}
for $\frac{\delta_p}{\sqrt 2 \Delta_p}\ll 1 $ , the orthogonal component up to a good approximation is 
\begin{equation}
|\chi^{\perp}\rangle_2  =\mathcal{N}\int_{-\infty}^{+\infty} p e^{-\frac{p^2}{2\Delta_p^2}}|p\rangle dp.
\end{equation}

For simplicity, we assume that the separation between $|L\rangle_1$ and $|R\rangle_1$ is sufficiently large so that we may, to a very good approximation, neglect the effect of the $|L\rangle_1$ component on mass-2 (the results hold even without this assumption). Thus, after the gravitational interaction between the masses their state will be
\begin{equation}
    \frac{1}{\sqrt{2}}(\sqrt{1+\epsilon}|L\rangle_1 |\chi\rangle_2+\sqrt{1-\epsilon}|R\rangle_1 |\chi'\rangle_2) 
    \label{shift}
\end{equation}
After postselection of state (\ref{final}) of mass-1, with the postselection probability being
\begin{equation}
P_{\text{Success}}=\frac{\epsilon^2}{4},
\end{equation}
the state of the mass-2 will be 
\begin{eqnarray}
   &&|\chi_{\rm post}\rangle_2 = \mathcal{N} (\sqrt{1+\epsilon} |\chi\rangle_2-\sqrt{1-\epsilon} |\chi'\rangle_2) \nonumber\\
 &\approx&( |\chi\rangle_2+ \frac{\delta_p}{\sqrt 2 \Delta_p ~\epsilon} |\chi^{\perp}\rangle_2) 
    \label{shift2}
\end{eqnarray}
The interaction is very small, so we can choose small $\epsilon $ such that not only $\frac{\delta_p}{\sqrt 2 \Delta_p }\ll 1$ holds, but also $\frac{\delta_p}{\sqrt 2 \Delta_p~ \epsilon}\ll 1$. Then (\ref{shift2}) corresponds to  amplified and reversed momentum obtained by mass-2 as was expected from the weak value of projection on arm $R$, $({\bf P}_R)_w=   - \frac{1}{\epsilon}$. Indeed, up to a very good approximation 
\begin{equation}
   |\chi_{\rm post}\rangle_2 =  \mathcal{N}\int_{-\infty}^{+\infty} e^{-\frac{(p-\frac{\delta_p}{\epsilon})^2}{2\Delta_p^2}}|p\rangle dp.
    \label{shift4}
\end{equation}

\begin{figure}
    \includegraphics[scale=0.26]{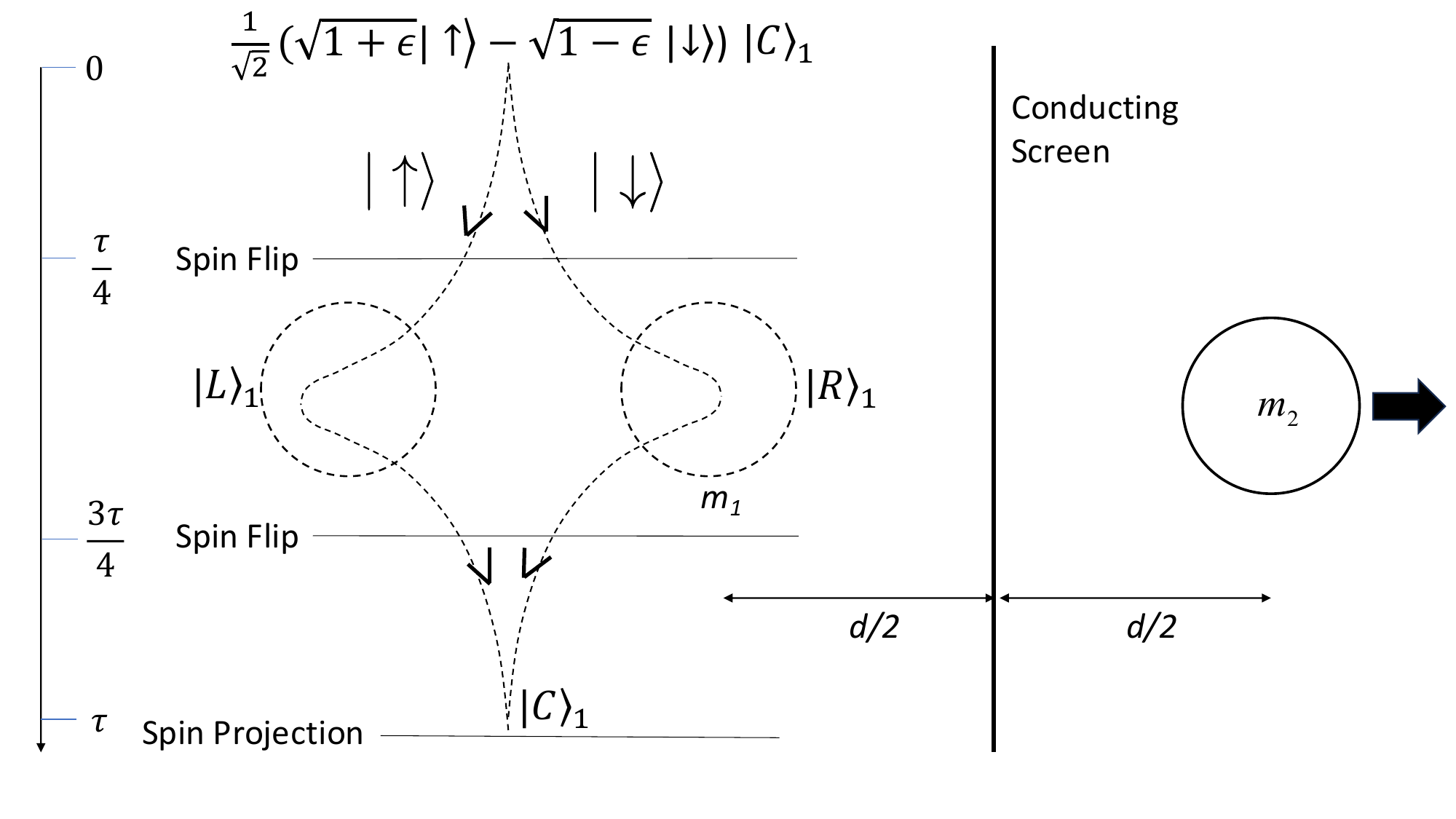}
    \caption{Mass-1 undergoing a Stern-Gerlach interferometry is outlined. The procedure uses an inhomogeneous magnetic field to accomplish spin-dependent splitting of paths. Two spin flipping pulses (also called $\pi$-pulses) are used to complete the interferometry as shown in the figure. The time axis is shown on the left. At the end of the interferometry, at time $\tau$, the spin is postselected in the $(|\uparrow\rangle_Q-|\downarrow\rangle_Q)/\sqrt{2}$ state, for which mass-2 has acquires a momentum away from mass-1.}
    \label{explicit}
\end{figure}

{\em Potential experimental scheme:}  One of the key ingredients is one of the masses, namely mass-1 in a quantum superposition of two distinct positions. We herewith plan to use an interferomtery of a large mass induced by a qubit \cite{bose1999scheme,armour2002entanglement,bose2006qubit}. One example is when the qubit is a spin embedded in a nanocrystal, which is called Stern-Gerlach interferometry, where a classic example can be a Nitrogen Vacancy (NV) centre spin in a diamond nanocrystal \cite{scala2013matter,wan2016free,margalit2021realization,marshman2022constructing}. An outline of a possible experimental scheme following this is given in Fig.\ref{explicit}. In this setting, the spatial superposition is created by spin dependent movements of the entire nanocrystal in a constant magnetic field gradient of $\frac{\partial B}{\partial x}$. Thus, the embedded spin (labeled by $Q$ for qubit) is initialized to $\frac{1}{\sqrt{2}}(\sqrt{1+\epsilon}|\uparrow\rangle_Q+\sqrt{1-\epsilon}|\downarrow\rangle_Q)$ and coupled to the center of mass of mass-1, say initially in the state $|C\rangle_1$, which stands for a Gaussian wavepacket at a certain location (such a state may be prepared by trapping and cooling). Then an evolution for a time $\tau/4$ in $\frac{\partial B}{\partial x}$, a spin flip via a microwave pulse, and a further evolution for a time $\tau/4$ in the {\em same} $\frac{\partial B}{\partial x}$, results in the following evolution
\begin{eqnarray}
\frac{1}{\sqrt{2}}(\sqrt{1+\epsilon}|\uparrow\rangle_Q+\sqrt{1-\epsilon}|\downarrow\rangle_Q)|C\rangle_1 \nonumber\\\rightarrow \frac{1}{\sqrt{2}}(\sqrt{1+\epsilon}|\uparrow\rangle_Q|L\rangle_1+\sqrt{1-\epsilon}|\downarrow\rangle_Q|R\rangle_1).
\label{spinev}
\end{eqnarray}
The right hand side of Eq.(\ref{spinev}) should then replace the state of mass-1 in Eq.(\ref{initial}), and its corresponding
part in (\ref{shift}), i.e., in Eq.(\ref{shift}), one should replace 
\begin{equation}
   |L\rangle_1\rightarrow |\uparrow\rangle_Q|L\rangle_1,~~~|R\rangle_1\rightarrow |\downarrow\rangle_Q |R\rangle_1. 
\end{equation} Subsequently, the superposition is undone (or closed) i.e., a full interferometry is completed on mass-1 by the inverse of the evolution depicted in Eq.(\ref{spinev}). This is accomplished by a further evolution for a time $\tau/4$ in the same $\frac{\partial B}{\partial x}$, a spin-flip pulse and then a final evolution for time $\tau/4$ in the same $\frac{\partial B}{\partial x}$, at the end of which both components are together. Thus the Stern-Grelach interferometry is completed in a time $\tau$. 
All the above is accomplished to ensure, at least in the ideal case, that there are no residual correlations between
the final state of the centre of mass of mass-1 and the states of the qubit embedded in it at the end of the interferometry. 

In order to have a prominent effect of mass-1 on mass-2 it is important for them to be as close as possible. So that they do not interact electromagnetically over short distances, a conducting/superconducting \cite{van2020quantum,schut2023micron} screen should be placed between the masses. We will assume that mass-2 is a trapped mass in a harmonic oscillator well of frequency $\omega$ initially cooled to its ground state. As soon as the superposition state of mass-1 starts being created, at say, time $\tau=0$, mass-2 is released from the trap to have free particle dynamics.

If this is all
ensured, then after the inverse evolution, the state of the qubit embedded in mass-1 and the centre of mass of mass-2 becomes
\begin{equation}
    \frac{1}{\sqrt{2}}(\sqrt{1+\epsilon}|\uparrow\rangle_Q |\chi\rangle_2+\sqrt{1-\epsilon}|\downarrow\rangle_Q |\chi'\rangle_2), 
    \label{shiftspins}
\end{equation}
so that in practice we measure and postselect on the outcome $\frac{1}{\sqrt{2}}(|\uparrow\rangle_Q-|\downarrow\rangle_Q)$, instead of $\frac{1}{\sqrt{2}}(|L\rangle_1-|R\rangle_1)$.

 For simplifying calculations and to get the order of magnitude effect, we will assume that as soon as mass-2 is released from its trap, it is subjected to the gravitational effect of mass-1 in a superposition of constant size, where the $|L\rangle_1$ and $|R\rangle_1$ states are separated by $\Delta x$. In practice, the superposition takes time to create and mass-2 will be released just when the superposition is starting to be prepared (i.e., the spin state dependent movement commences) and let it evolve under gravitational interaction for the full duration that mass-1 is in superposition: one only has to replace the full calculation with a $\Delta x(t)$, which will still give the same order of magnitude movements of mass-2 as when it is subjected to a superposition of fixed size $\Delta x$. With the parameters that we will choose here, our approximation of negligible effect of mass-1 on mass-2 for the initial state $|L\rangle_1|\chi\rangle_2$ is valid 
(strictly speaking, this is not prohibitive, as for $\Delta x< d$, we will simply look for an appropriate relative momentum shift, although that will be suppressed by an extra factor of $\frac{\Delta x}{d}$ in comparison to the momentum shift we will get for our parameters). In this case,
the gravitational force acting on mass-2 is $F=\frac{Gm_1 m_2}{d^2}$, so that the momentum transferred during the interaction is 
\begin{equation}
\delta_p \sim F\tau= \frac{Gm_1 m_2}{d^2}\tau,
\label{deltap}
\end{equation}
where $m_1$ and $m_2$ depict the magnitudes of the masses of mass-1 and mass-2 respectively.
Conditional on the outcome $\frac{1}{\sqrt{2}}(|\uparrow\rangle_Q-|\downarrow\rangle_Q)$ when measuring the qubit, this momentum becomes $-\frac{\delta_p}{\epsilon}$, i.e., both amplified and converted to be opposite to $\delta_p$, i.e., repulsive. Measuring this momentum change is thus the evidence of the superposition of gravitational fields. 
This momentum is measured, in the end, by {\em measuring the position} $x_2(T) \sim -\frac{\delta_p}{\epsilon} \frac{T}{m_2} $ of $m_2$ at a given long time, say $T >> \tau$, after release, during which the wavefunction also expands as $\Delta_p \frac{T}{m_2}$. Thus, the two relevant entities to compare are $\frac{\delta_p}{\epsilon}$ and $\Delta_p$ as they both get multiplied by the same factor $\frac{T}{m_2}$. We thus simply need to compare two Gaussians in momentum -- the original wavepacket of mass-2 ($|\chi\rangle_2$) and the momentum shifted wavepacket ($|\chi_{\rm post}\rangle_2$). Thus, as we need to have a statistically significant discrimination of Gaussian wavepackets of width $\Delta_p$ whose peaks are separated by $\frac{\delta_p}{\epsilon}$ while still maintaining $\frac{\delta_p}{\epsilon \sqrt{2}\Delta_p}<1$, it will be desirable to get
\begin{equation}
\frac{\delta_p}{\epsilon} \sim 0.1~\Delta_p,
\label{cond}
\end{equation}
by appropriately choosing $\epsilon$ while at the same time ensuring that the probability of the right postselection $\frac{\epsilon^2}{4}$ does not get too small.
The preparation of the initial wavepacket via cooling in a harmonic trap depends on $m_2$ as $\Delta_p = \sqrt{\frac{m_2\omega \hbar}{2}}$, while with an extra momentum squeezing by a factor $\eta<1$, possible by periodically varying the frequency of the trap after cooling to the ground state, we can get $\Delta_p = \eta \sqrt{\frac{m_2\omega \hbar}{2}}$.  Using this $\Delta_p$ and Eq.(\ref{deltap}) for $\delta_p$ in our desirable condition Eq.(\ref{cond}), we get
\begin{equation}
\epsilon \sim  \frac{10~Gm_1\sqrt{\frac{2m_2 }{ \omega \hbar}} \tau}{\eta~d^2}.
\label{cond-mass}
\end{equation}

{\em Parameters:}. We will envisage a separation of $d\sim 20 \mu m$ between the closest approach of the masses, i.e., the situation
$|R\rangle_1|\chi'\rangle_2$, so that the conducting/superconducting screen (which will itself be a few microns thick) and some separation of the each of the masses from the screen can be accommodated \cite{van2020quantum,schut2023micron,schut2023relaxation} so as to not be overwhelmed by surface forces. Roughly, we assume $\sim 10~\mu$m separation of the masses on either side of the screen. Note that this immediately restricts the value of $m_2$ to be no larger than $\sim 10^{-12}$ kg as the radius of such particles (say, for standard densities of solids such as diamond) is already $\sim 4 \mu$m. We will examine the scheme  for $m_2 \sim 10^{-12}$ kg.

  Next is the question of the values of the $m_1$ and the superposition size $\Delta x$. For $m_1 \sim 10^{-14}$ kg can assume that superposition sizes of $\Delta x \geq 50 \mu$m will be possible by Stern-Gerlach interferometry -- with  sufficiently strong but feasible $\frac{\partial B}{\partial x}$ and assuming a coherence time long enough to allow for the $\tau \sim 1$s time-scale of the interferometry. Creation of such large superpositions via the Stern-Gerlach effect has been extensively discussed \cite{bose2017spin,marshman2020mesoscopic,marshman2022constructing,zhou2022catapulting,zhou2022mass,braccini2024exponential,zhou2025spin} (presumably, spinless methods can also be adapted \cite{romero2011large}, as well as spinless methodologies for a measurement that projects to $\frac{1}{\sqrt{2}}(|L\rangle-|R\rangle)$ as an outcome \cite{yi2021massive}). The conditions needed to ensure the coherence of such large superpositions for sufficiently long times (e.g, $\tau \sim 1$ s) have also been worked out, giving, for example, relevant values for pressure \cite{romero2011large,bose2017spin,van2020quantum,schut2025expression}, temperature \cite{romero2011large}, electromagnetic control and noise \cite{narasimha2025magnetic}, and inertial and gravitational noise \cite{torovs2020relative}. These conditions will also have to be met. At the moment, we cannot envisage a method to produce similar superpositions for a larger mass in $\tau \sim 1$s using Stern-Gerlach interferometry techniques. 
  
Putting all the above parameter values in Eq.(\ref{cond-mass}), and assuming a $\omega \sim 1$ (possible in diamagnetic traps created by current wires on chips \cite{elahi2025diamagnetic}, so that one may release the mass as  needed by switching fields off), we get
$
\epsilon \sim  \frac{3.5\times 10^{-3}}{\eta}$.
Thus, choosing a momentum squeezing of $\eta \sim 0.35$ should make it feasible to make $\epsilon \sim 0.01$. We may want higher $\omega$ as cooling to ground state \cite{delic2020cooling,tebbenjohanns2021quantum,magrini2021real,piotrowski2023simultaneous} for nanocrystals has to date been achieved in substantially higher frequency traps. One may also want to use smaller masses $m_2$ in the experiments. According to Eq.(\ref{cond-mass}), an increase in $\omega$ or a decrease in $m_2$ can be adjusted by the appropriate squeezing parameter $\eta$, especially since ground state squeezing has also been realized for trapped nanocrystals \cite{kamba2025quantum}. For example, if we want to use $m_2 \sim 10^{-14}$ kg (micrometer-sized) nanocrystals and $\omega \sim 10$ kHz (diamagnetic traps where it should be feasible to prepare the ground state \cite{elahi2025diamagnetic}), we will adjust the momentum squeezing parameter of $\eta \sim 10^{-3}$ to keep $\epsilon \sim 10^{-2}$.


 For the above values, we estimate $\epsilon \sim 0.01$, so that only a fraction of $\sim \frac{\epsilon^2}{4}\sim 2.5\times 10^{-5}$ of the outcomes are retained. Under this retention, one has to detect the positions of nanoparticles with a resolution finer than $\frac{\delta_p}{\epsilon} \frac{T}{m_2} \sim 1 \AA$ taking, say, $T\sim 10$s which is feasible \cite{whittle2021approaching,gieseler2012subkelvin,vovrosh2017parametric}. Even with this capability, a further task through these momentum measurements is the discrimination of the momentum shifted and unshifted Gaussians, which happens with the probability $1-|\langle \chi|\chi_{\rm post}\rangle|^2 \approx (\frac{\delta_p}{\epsilon \Delta_p})^2 \sim 10^{-2}$. Thus the total probability of successfully detecting the movement of $m_2$ away from $m_1$ is $\sim 2.5\times 10^{-7}$, which implies $\sim 4 \times 10^7$ runs of the experiment.

 {\em Conclusions:} Here, we have presented a proposal for a table-top experiment demonstrating the quantumness of the gravitational interaction. It implements a weak measurement amplification scheme, which is a quantum interference effect, so, if successful, will represent direct empirical evidence of quantum nature of gravity. A brief analysis of the feasibility of the experiment shows that it is on the verge of current technology.
 
 {\em Addendum:} On the completion of this work, we became aware that another independent work reporting a similar idea has recently appeared \cite{saldanha2026repulsive}. 
\begin{acknowledgments}
{\it Acknowledgements:--}  This work was conceived during a Mathematical and Physical Sciences Visiting Fellowship of LV at University College London in 2024. SB would like to acknowledge EPSRC grant EP/X009467/1 and STFC grant ST/W006227/1. This work was made possible through the
support of the WOST, WithOut SpaceTime project (https://withoutspacetime.org), supported by Grant ID\#63683
from the John Templeton Foundation (JTF). S.B’s research is funded by the Gordon and Betty
Moore Foundation through Grant GBMF12328, DOI
10.37807/GBMF12328, and the Alfred P. Sloan Foundation under Grant No. G-2023-21130. LV’s research is supported in part by the Israel Science Foundation Grant  No.~2689/23
\end{acknowledgments}

\bibliography{VB} 

\end{document}